\documentclass{sig-alternate}
\usepackage[utf8]{inputenc}
\usepackage{url}
\usepackage{units}

\begin{document}

\title{Development and Testing of Automotive Ethernet-Networks together in one Tool - OMNeT++ 
}
%
%
%
%
%

\numberofauthors{2} 
%
\author{
%
%
\alignauthor
Patrick Wunner \\
       \affaddr{iSyst GmbH}\\
       \affaddr{Nordostpark 91}\\
       \affaddr{Nürnberg, Germany}\\
       \email{patrick.wunner@isyst.de}
\alignauthor
Stefan May\\ 
       \affaddr{TH Nürnberg}\\
       \affaddr{Keßlerplatz 12}\\
       \affaddr{Nürnberg, Germany}\\
       \email{stefan.may@th-nuernberg.de}
\and  
\alignauthor Kristian Trenkel \\ 
       \affaddr{iSyst GmbH}\\
       \affaddr{Nordostpark 91}\\
       \affaddr{Nürnberg, Germany}\\
       \email{kristian.trenkel@isyst.de}
\alignauthor Sebastian Dengler\\
       \affaddr{iSyst GmbH}\\
       \affaddr{Nordostpark 91}\\
       \affaddr{Nürnberg, Germany}\\
       \email{sebastian.dengler@isyst.de}
}

\maketitle
\begin{abstract}

In this paper, the network simulation framework OMNeT++ is used for development and testing of automotive Ethernet-Networks. Therefore OMNeT++ is extended by the INET framework. It is augmented by an implementation of the protocol SOME/IP (-SD) and an connector to the middleware Gamma V. The middleware is used to configure the network by initialization. Additionally data, which is sent over the network, can be changed on the fly. \\
The contribution of this work regards three main aspects: First, the use of OMNeT++ for network development in automotive industry. Second, the employment of an existing simulation model and using it as restbus simulation for Hardware in the Loop (HiL) testing or rapid prototyping. Finally, the implementation of SOME/IP(-SD) into OMNeT++.
\end{abstract}





\section{Introduction}
More and more electrical components in the automotive industry are communicating no longer by classic fieldbuses, but by real-time Ethernet bus systems. Bus systems such as CAN, FlexRay or MOST used up to now are replaced by real-time Ethernet protocols little by little: TTEthernet, Audio-Video-Bridging (AVB), Time Sensitive Networking (TSN), etc. This evolution is triggered by the constantly increasing necessity of bandwidth for data transfer and the modular and flexible architecture of Ethernet. In consequence a new physical layer, BroadR-Reach \cite{_broadr-reach_2014}, replaces the standard layer. Due to that, a bidirectional communication of \unit[100]{MBit/s} can be established by one unshielded twisted pair of cable.\\
But for these benefits the complexity and the test investments of the network increase. Because of this, OMNeT++ is used for pure and restbus simulation with the same model.\\
To show the functional use of this solution the Scalable service Oriented MiddlewarE over IP (SOME/IP) with Service Discovery (SOME/IP-SD) was implemented into OMNeT++. This protocol was developed by BMW and has been integrated in the AUTOSAR 4.1 specifications. It is based on TCP/UDP transport layer. In consequence the INET framework is taken to build up the underlaying protocol layers IP and TCP/UDP. A simulation model has been created, that can be run as pure or restbus simulation. The reaction time of the simulated hosts and the maximal workload were analyzed for validation. For communication to real BroadR-Reach networks a converter between the automotive physical layer and the standard can be taken. 

\subsection{Related Work}
There are several Ethernet-Test solutions available on the market that can be used in a HiL environment. The dSPACE Ethernet Blockset \cite{_dspace_2014} is mainly used for unit and function test. The Vector CANoe.Ethernet \cite{_canoe.ethernet_2014} is a standalone solution which can be controlled from a control desk to expand the HiL tests. For testing existing nodes they work fine, but for developing whole systems or ECUs they are not as flexible as needed.\\
The Communication over Real-time Ethernet (CoRE) research group Hamburg University of Applied Sciences \cite{_core4inet_2014} uses the OMNeT++ Framework for simulation. The focus of their work is certainly the pure simulation of Ethernet networks. Therefore the real-time protocols TTEthernet \cite{_ttethernet:_2014} and IEEE 802.1 AVB, which is renamed to TSN \cite{_ieee_2013}, were implemented into OMNeT++.\\
In this work OMNeT++ is used for restbus simulation. A pure simulated network can be taken to replace single simulated hosts by real hardware. So it can be used for developing network communication and ECUs. 


\section{Development and Testing of automotive Ethernet-Networks}
A lot of different companies and engineers are involved by building up an in-vehicle network. In the next sections problems (Section \ref{cha:Development and Testing of automotive Ethernet-Network:Problems_in_Testing}) and development solutions (Section \ref{cha:Development and Testing of automotive Ethernet-Network:Pure Simulation} - \ref{cha:Development and Testing of automotive Ethernet-Network:Restbus Simulation}) are shown. Additionally the protocol SOME/IP (-SD) is explained (Section \ref{cha:Development and Testing of automotive Ethernet-Network:SOME_IP}).

\subsection{Problems in Testing}
\label{cha:Development and Testing of automotive Ethernet-Network:Problems_in_Testing}
\begin{figure}[h]
\centering
\includegraphics[width=0.9\linewidth]{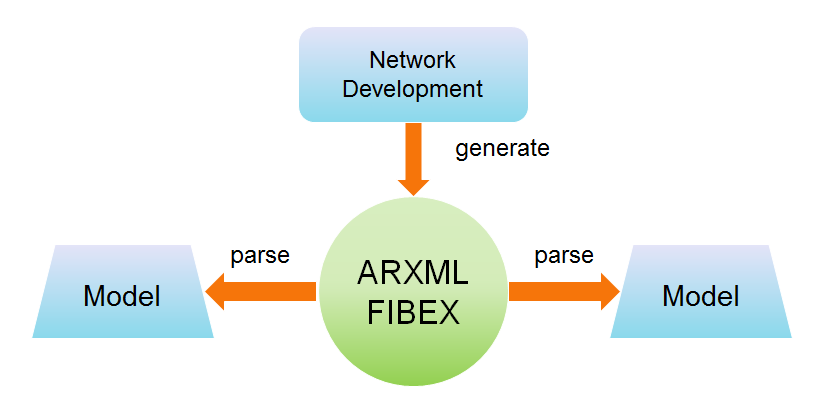}
\caption{Model description files for model generation}
\label{fig:FIBEX}
\end{figure}

For a single in-vehicle network many simulation models have to be created. The automotive industry tries to address this problem by consistent bus description files, such as Fieldbus exchange format (FIBEX) or AUTOSAR XML (ARXML). In principle with these files it is possible to generate a simulation model automatically (Figure \ref{fig:FIBEX}). Unfortunately tools interpret data differently, so an automatic test generation often does not work and generated models have to be reconfigured for the special project.

The concept in this paper is different. Instead of creating several models, OMNeT++ is chosen to fulfill all these requirements. A pure simulation model can be created for conceptual design. With pcapRecorders the traffic at all hosts can be traced and afterwards analyzed by standard tools e.g. wireshark. The structure of the network and the bandwidth of every connection can be changed easily. After that simulated ECUs or branches of the network can be replaced with real hardware step by step. Therefore tracepoints in the simulation can be kept. Engineers and testers can hereby test, if the ECUs work as specified. Single network nodes may be replaced by a real ECU, while the restbus keeps resting upon the simulation. Thus ECUs can be tested against the same model.

\subsection{Pure Simulation}
\label{cha:Development and Testing of automotive Ethernet-Network:Pure Simulation}
At first a board net engineer has to design the network. Therefore several aspects have to be considered:
\begin{itemize}
\item Structure of the network
\item Number of switches and nodes
\item Data to be transmitted
\item Metrics for prioritization
\end{itemize}

\begin{figure}[h]
\centering
\includegraphics[width=0.9\linewidth]{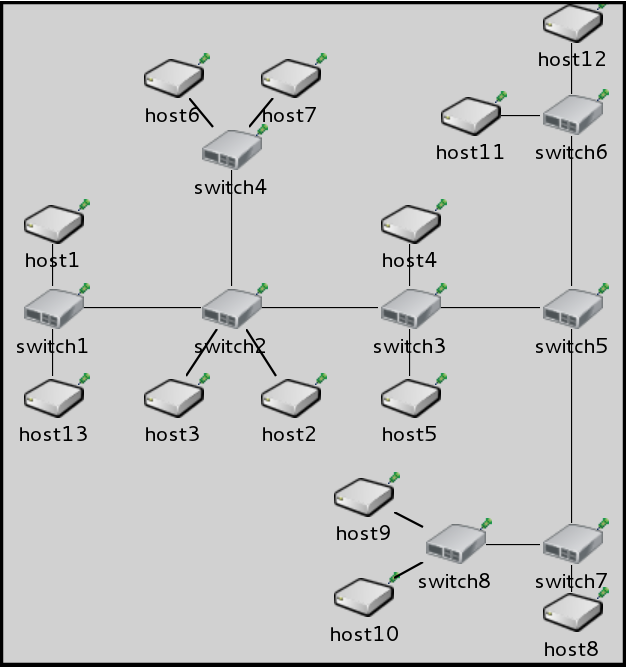}
\caption{Pure network simulation in OMNeT++}
\label{fig:complexNetwork}
\end{figure}

In a shared bus system, such as LIN or CAN, the bus load is the same at every physical location. In switched networks, in contrast, the load can variate between the single branches. With the use of switched Ethernet every node has a \unit[100]{MBit/s} point-to-point connection to a switch. This is an enormous advantage compared to shared bus systems, where the bandwidth has to be split between the nodes. If the switched network is well designed, every node can send and receive data exploiting the whole \unit[100]{MBit/s} bandwidth. Additionally a \unit[1]{GBit/s} connection between a host and a switch on the same board can be realized to increase the efficiency of the network. For high load of the whole network it is necessary to find bottlenecks in a complex system of switches and Electronic Control Units (ECUs). Figure \ref{fig:complexNetwork} shows such a network that is build up in OMNeT++. Finding bottlenecks in a real network is hard to realize. Special switches with monitoring ports or hubs, which are connected to network analysis components, have to be integrated in several branches. With simulation tools, such as OMNeT++, many trace points and several tests can be realized to scale the network to its best performance with little effort.
\subsection{Restbus Simulation}
\label{cha:Development and Testing of automotive Ethernet-Network:Restbus Simulation}
The term "Restbus Simulation" is used in the automotive industry. It describes the simulation of the bus communication of one or more bus subscriber. The in-vehicle communication is based on bus systems like CAN, FlexRay or Ethernet. The development of the single bus subscribers or ECUs is done by different suppliers. So for the development of one of the ECUs the corresponding nodes are not available. Usually the closed-loop control functionality is distributed on different nodes on the bus. Due to that, most of the ECUs functionality does not work autonomously. Therefore it is necessary to simulate or emulate the other bus subscriber.\\
In the context of this paper, the restbus simulation is a  further development of the pure simulation described in section \ref{cha:Development and Testing of automotive Ethernet-Network:Pure Simulation}. The next sections explain the usage of the restbus simulation for the development as rapid prototyping (\ref{cha:Development and Testing of automotive Ethernet-Network:Restbus Simulation:Rapid Prototyping}) and for the testing as Hardware in the Loop (\ref{cha:Development and Testing of automotive Ethernet-Network:Restbus Simulation:Hardware in the Loop}).

\begin{figure*}
\centering
\includegraphics[width=0.7\linewidth]{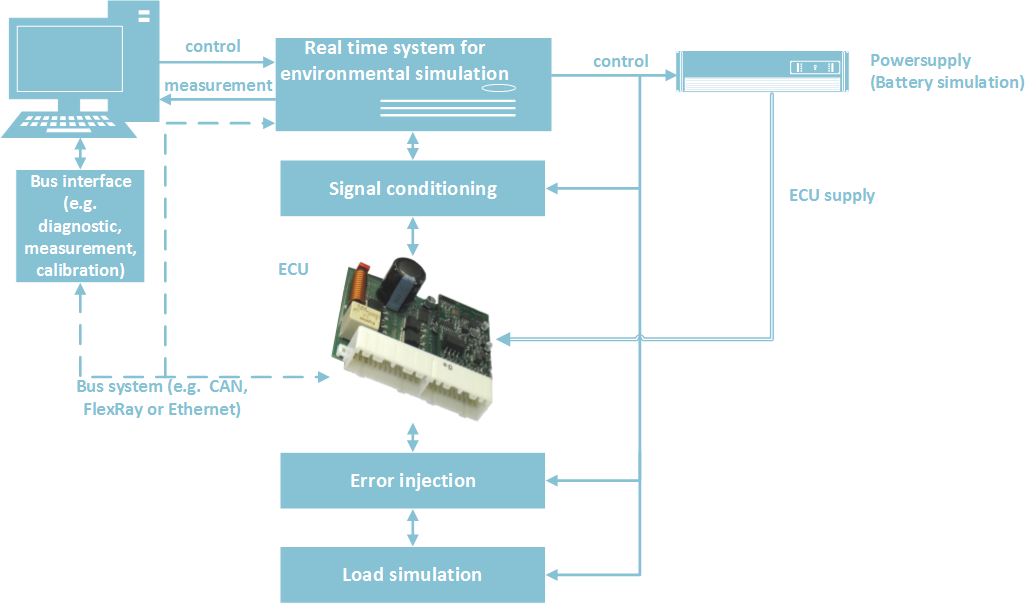}
\caption{Structure of a typical HIL system}
\label{fig:HIL_System}
\end{figure*}

\subsubsection{Rapid Prototyping}
\label{cha:Development and Testing of automotive Ethernet-Network:Restbus Simulation:Rapid Prototyping}
Rapid prototyping is a kind of development process, which is used in the early stages of development in the automotive industries. 
The aim of this approach is to check the feasibility of a functional prototype. In this stage no special hardware for the ECU is available. Sometimes the hardware of an older ECU is used, but most of the prototypes are realized on real time hardware like the dSPACE micro autobox \cite{_dspace_2014-1} or PC hardware. For the prototype only the main functionality of the planed system is implemented.\\
Based on the given process two different use cases for restbus simulation for the rapid prototyping can be distinguished.\\
The first usage is the simulation of the other bus subscribers during the development. For the implementation of distributed control functionality it is necessary to simulate the other involved ECUs. For example the ECU for an active damper control system needs values of the height sensors of the wheels from one or more separate ECUs. Additionally it needs the acceleration and gear information provided from an other ECU on the bus system. For the development of such an active damper control system at least the simulation of these sensor values is needed.\\
The second usage is testing a prototype car. For this it is necessary to simulate one or more ECUs that are not realized yet. So you can send single frames on the bus of an existing system.
The solution realized in this work can be used for both above mentioned cases of the restbus simulation for rapid prototyping. \\
Based on the pure simulation of a network one or more nodes can be disconnected from the simulation and a real ECU can be connected. With a router inside the simulation framework the connection to a real network interface can be established. Not only whole ECUs but also single services of SOME/IP can be simulated. \\
Through the integration with the real-time middleware Gamma V (for more information see section \ref{cha:Implementation:Connection to Middleware (GammaV)}) the framework can also be used as complete rapid prototyping system.\\

\subsubsection{Hardware in the Loop}
\label{cha:Development and Testing of automotive Ethernet-Network:Restbus Simulation:Hardware in the Loop}
Hardware in the Loop (HiL) is a test concept mostly used in the automotive industries. But it is also used in the branches aerospace and industrial. The concept of HiL bases on the connection of one ECU to a simulation system which simulates the whole environment of the ECU (e.g. analog and digital sensors and actors, bus systems and further more). The structure of a typical HiL system is shown in figure \ref{fig:HIL_System}. It demonstrates the single components, that are needed in a HiL environment. The aim of the HiL test is the system test according to the requirements. Therefore an accurate and real-time capable simulation of the environment is needed. The real-time requirement for cycle execution of the model in automobile systems is typically 1\,ms.\\
One major part of the environmental simulation is the restbus simulation of the bus system. The simulation consists of the correct frames, timings and content on the bus. In cooperation with behavior models (e.g. Matlab/Simulink models) of other ECUs the content is calculated. For the testing purpose the simulation of errors (e.g. wrong values, checksum, etc.) is needed, too.\\
The restbus simulation for the HiL testing can also be based on the pure simulation model. Like in rapid prototyping approach one or more nodes can be deleted from the simulation and a connection to the real world can be realized.\\
Depending on the used real-time system the developed framework can be used in two different ways.\\
On the one hand a separate PC can be used for restbus simulation. This is needed in case the real-time systems are not open for own extensions or that it is not running Linux as operating system (e.g. dSAPCE real time hardware). \\
On the other hand in open real-time systems based on Linux the framework can be integrated in the real time system. In that case a fully integrated restbus simulation can be realized. A direct integration to the simulation models of the ECUs is also possible.\\\\\\

\subsection{The Protocol SOME/IP (-SD)}
\label{cha:Development and Testing of automotive Ethernet-Network:SOME_IP}
SOME/IP is a UDP/TCP-based network protocol that has been developed by BMW in scope of a promoted project. It is the only solution which complies with all automotive requirements and is compatible to AUTOSAR\cite{_autosar_2013}. The advantage in comparison to other protocols is that needed data is only sent from the host to the client when the client is subscribed to the service.
\begin{figure}[h]
\centering
\includegraphics[width=0.9\linewidth]{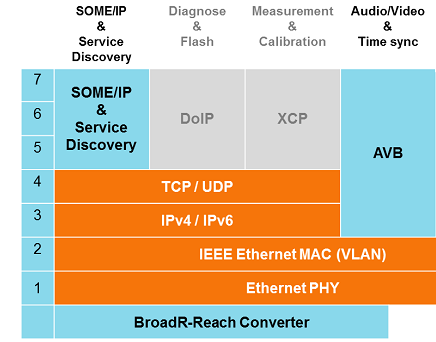}
\caption{OSI model and classification of protocols}
\label{fig:ISO-OSI}
\end{figure}
For communication some standard components from the Ethernet stack are used. In figure \ref{fig:ISO-OSI} the location of the SOME/IP (-SD) protocol in the OSI model is shown. IP and TCP/IP layers are used unmodified. The MAC layer is extended by a Virtual LAN (VLAN) tag. The physical layer in contrast has been substituted. Instead of the shielded CAT cable with four or eight wires the BroadR-Reach technology of Broadcom is used. \\
The advantage of this solution is a bidirectional connection with the bandwidth of \unit[100]{MBit/s} over an unshielded drilled cable pair.\\
SOME/IP (-SD) is not a real time protocol because it is only based on standard Ethernet layers. Actually the real-time requirements are met by a moderate covered Quality of Service (QoS). For hard real-time requirements the AVB protocol is intended.\\
The layers for diagnostic and flash, such as measurement and calibration, are not explicated in this work.

\section{Implementation}
In the following sections the implementation of SOME/IP(-SD) (Section \ref{cha:Implementation:SOME_IP}) and the middleware Gamma V (Section \ref{cha:Implementation:Connection to Middleware (GammaV)}) is described.
\subsection{Implementation SOME/IP (-SD)}
\label{cha:Implementation:SOME_IP}
\begin{figure}[h]
\centering
\includegraphics[width=0.9\linewidth]{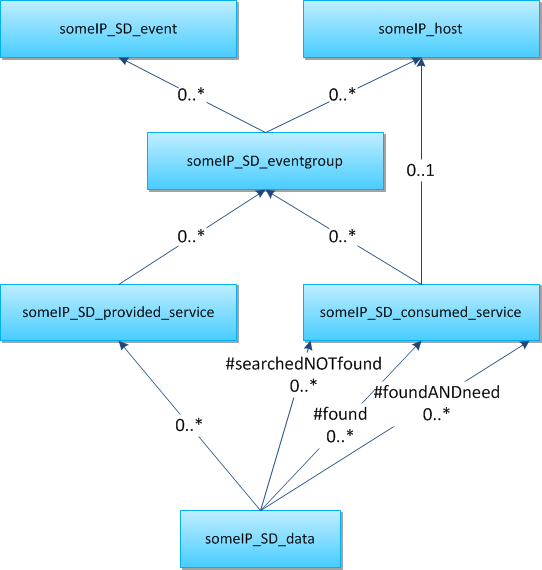}
\caption{UML Diagram of the data model}
\label{fig:sd_data_uebersicht}
\end{figure}

For SOME/IP (-SD) implementation the INET Framework is extended. This framework offers several standard Ethernet protocols and layers such as IP, UDP, TCP, etc. that are basically used by the in-vehicle network SOME/IP (-SD). Additionally it provides the link to the real network card of the simulation host. So a restbus simulation can be built up.\\
The characteristic of all simulated SOME/IP nodes is different. That is the reason for the service oriented data transfer. Every node needs and offers some services. The Service Discovery (SD) protocol is implemented in SOME/IP to find and subscribe services. \\
A data model is implemented to store these information. Figure \ref{fig:sd_data_uebersicht} shows a diagram about the data model. First, the simulation has to be connected to the middleware Gamma V. Afterwards the data models for every simulated node have to be filled with information in the initialization phase of the simulation. All services that are provided and consumed including their events and eventgroups are read out. Based on this information the dynamic internal data model is generated.\\
To realize the current behavior of what a node has to send over the network, it has to know the following things:

\begin{itemize}

\item What services are needed? \\(\texttt{someIP\_SD\_consumed\_service})

\item What services have to be offered? \\(\texttt{someIP\_SD\_provided\_service})

\item Which nodes are subscribed to my services? \\(\texttt{someIP\_host})

\item What eventgroups and events contain a service?
\\(\texttt{someIP\_SD\_eventgroup, someIP\_SD\_event})

\item Which nodes offers needed services? \\(\texttt{someIP\_host})

\item To which services am I subscribed?

\item When was the last event on a service? Is the "Time to live" exhausted?

\item Which data is included in the payload of offered and needed services?

\end{itemize} 

\subsection{Connection to Middleware Gamma V}
\label{cha:Implementation:Connection to Middleware (GammaV)}

\begin{figure*}
\centering
\includegraphics[width=0.7\linewidth]{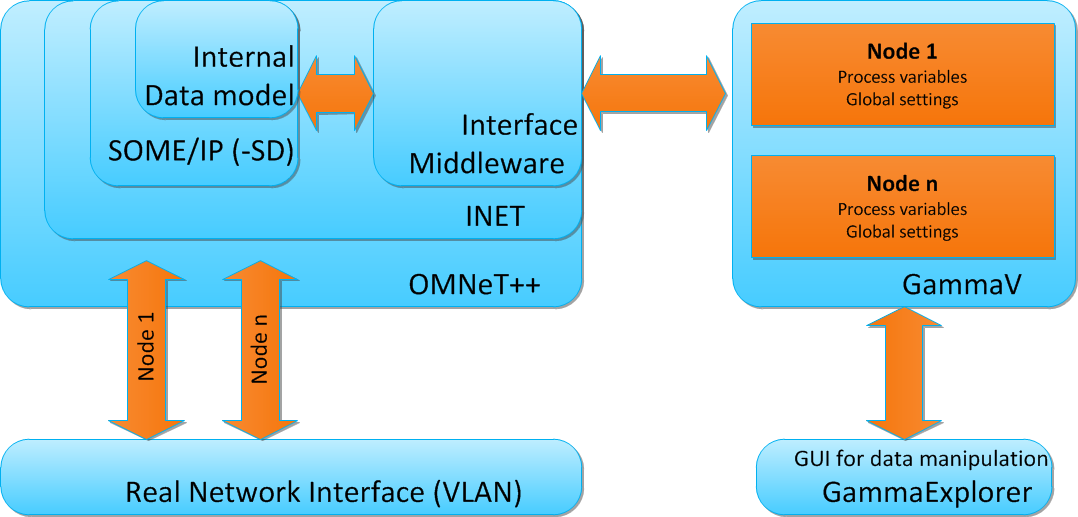}
\caption{Middleware Integration Overview}
\label{fig:overview}
\end{figure*}
\begin{figure}[h]
\centering
\includegraphics[width=0.9\linewidth]{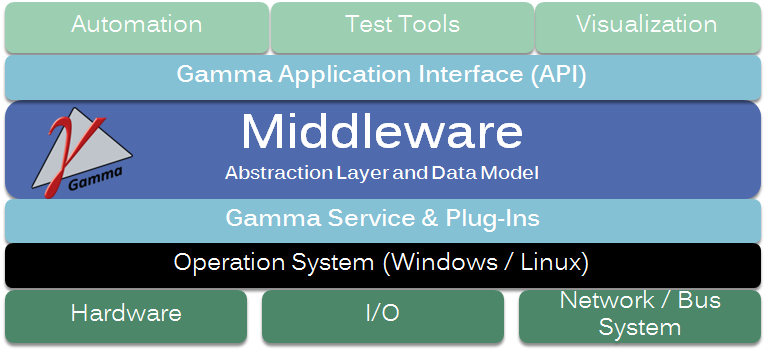}
\caption{Structure of the Gamma V Middleware}
\label{fig:Gamma_V_Structure}
\end{figure}

The software Gamma V is a middleware system provided by the company RST \cite{_gamma_2014}. A schematical overview can be found in figure \ref{fig:Gamma_V_Structure}. Like other middleware systems Gamma V provides an abstraction between the application software on the one side and the hardware and operation system on the other side. The interface to the application is realized by the Gamma API and consists of structured variables with different data types. The interface to the hardware and bus systems is realized by so called IO-Plug-ins. These plug-ins are small wrappers for the hardware driver and the Gamma V middleware.\\
In addition to this functionality the middleware also provides a timing model for the real time execution of applications. With this functionality it is possible to realize real-time environmental simulation as it is needed for the HiL testing. The Gamma API also allows the integration of Matlab/Simulink models by using a special blockset. By this integration the realization of environmental models for testing or the function development for rapid prototyping is possible.

Gamma V is used because of the following aspects:
\begin{itemize}
\item Support of the operating system Linux
\item Usability for rapid prototyping
\item Usability for HiL testing
\item Full integrated solution for development and testing
\end{itemize}
There are two possible solutions for the integration between the simulation framework and the middleware. The first solution is the implementation of an IO-Plug-in. This solution provides a fast access to the data model during runtime of the middleware but the effort of implementation is high. The second solution was the usage of the Gamma API. This way provides a slower access to the data model. But therefore the implementation is fast. An other advantage is that the simulation framework stays widely independent from the middleware.\\
For the current work the second solution is implemented. For most use cases the timing of this implementation is adequate. Figure \ref{fig:overview} shows an overview over the system structure. The internal data models of the SOME/IP implementation get their content from the middleware.\\
The data model is described in an XML file ("gamma.xml"). Therein the single hosts with their settings, the services, eventgroups, events and data fields are characterized. This file is used to build up the Gamma V model.

The middleware integration provides the following advantages:
\begin{itemize}
\item Central definition of the data structure for SOME/IP
\item Direct integration of the restbus simulation for rapid prototyping and HiL testing
\item Visualization of the signal values 
\end{itemize}

In initializing phase of the simulation the connection to the middleware is established. After that services, eventgroups and events of each host are read-out. With these information the internal data model is built up dynamically. To get the data fields, structures or arrays of the single events, an XML parser has been developed. This parser reads out the gamma.xml to generate a map between the data fields in the middleware and the internal data model. Therefore handles to the Gamma V fields are opened and stored.\\
While the simulation is running a parser and a serializer manages the communication between both data models when data is sent or received over the network (simulated or real). \\


\section{Experiments and Results}

The communication between the restbus simulation and the real ECU can be established. The services can be found bidirectionally over multicast addresses. The subscription over singlecast addresses is working. The time simulated hosts need to answer a frame is within the specification of actual automotive communication requirements.\\
The following goals could be realized until now:

\begin{itemize}

\item Pure simulation of in-vehicle SOME/IP(-SD) network

\item Create restbus simulation from simulation model

\item Subscribe, Unsubscribe, Resubscribe, Start, Stop Restart services

\item Connection to middleware Gamma V

\item Create internal data model in initialization phase

\item Manipulate values of data fields over middleware

\end{itemize}
The following graphics show the timing characteristics of the restbus simulation. The values are taken on an Intel Core i7-3520M with Linux Debian (Kernel 3.2.0-4-amd64) as operating system. The simulation has been compiled as release version and the \texttt{nice} value of the process has been set to -10.\\
%
\vspace{-10pt}
\begin{figure}[h]
\includegraphics{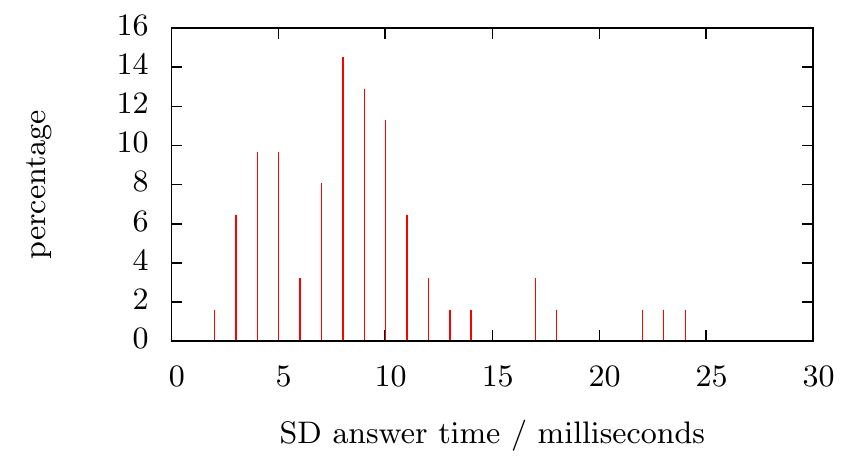}
\vspace{-20pt}
\caption{Answer time of SD messages}
\label{fig:sd_answer_time}
\end{figure}

%

\begin{figure}[h]
\includegraphics{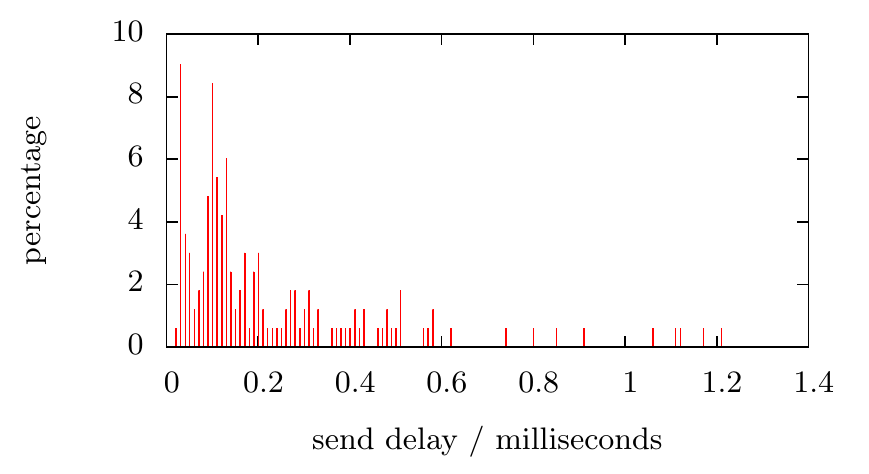}
\vspace{-20pt}
\caption{Send delay of messages}
\label{fig:send_delay}
\end{figure}

Figure \ref{fig:sd_answer_time} shows the time that is required for answering a SOME/IP Service Discovery request from an external source.\\
In figure \ref{fig:send_delay} the time is shown that is needed to send consecutive messages.
\section{Conclusions and Future Work}
The presented concept to use a network simulation framework for restbus simulation of Ethernet based connections fulfills all requirements. The modular structure and the strict separation of network description and protocol layer make it possible to adjust an existing simulation model to new requirements without C++ programming knowledge. Most settings can be done graphically or by the domain specific programming language NED.

To develop a full functional test system with the network simulation framework OMNeT++ or the commercial version OMNEST \cite{_omnest_2014}, some additional functions have to be researched and implemented:

\begin{itemize}

\item Research in other protocols with hard real time requirements

\item Automatic test build from a network description file

\item Error injection
\end{itemize}


%
\bibliographystyle{IEEEtran}
\bibliography{IEEEabrv,bibliothek2}
\end{document}